\documentclass[a4paper]{article}

\usepackage{INTERSPEECH2020}
\usepackage{booktabs}
\usepackage[export]{adjustbox}
\usepackage{amsmath,graphicx,bm}
\usepackage{multirow}
\usepackage{subfigure}
\usepackage{cite}

\title{Domain Aware Training for Far-field Small-footprint Keyword Spotting}
\name{Haiwei Wu$^{1}$, Yan Jia$^{1}$, Yuanfei Nie$^2$, Ming Li$^1$}
\address{
    $^1$Data Science Research Center, Duke Kunshan University, Kunshan, China\\
    $^2$Montage Technology, Kunshan, China}
\email{ming.li369@dukekunshan.edu.cn}

\begin{document}
    
    \maketitle
    \begin{abstract}
        
        In this paper, we focus on the task of small-footprint keyword spotting under the far-field scenario. Far-field environments are commonly encountered in real-life speech applications, causing severe degradation of performance due to room reverberation and various kinds of noises. Our baseline system is built on the convolutional neural network trained with pooled data of both far-field and close-talking speech. To cope with the distortions, we develop three domain aware training systems, including the domain embedding system, the deep CORAL system, and the multi-task learning system. These methods incorporate domain knowledge into network training and improve the performance of the keyword classifier on far-field conditions. Experimental results show that our proposed methods manage to maintain the performance on the close-talking speech and achieve significant improvement on the far-field test set.

    \end{abstract}
    
    
    \section{Introduction}
    
    Small footprint keyword spotting (KWS), also known as wake-up word detection, is a task to detect the occurrences of a pre-defined keyword in continuous speech signals. With the rapid development of mobile devices, smart speakers, and other applications, which require a hands-free conversational interface, this technology is attracting more and more attention. Different from the traditional keyword spotting task, with the constraints of hardware, real-life wake-up word detection must have a small memory and low computational cost. And simultaneously, it also requires to be highly accurate in detection and robust in different complex environments like noisy and far-field conditions.
    
    Traditional approaches \cite{chen2014small,  sun2017compressed} on this task involve Hidden Markov Models (HMMs), which are utilized to construct the keyword model and the filler/background model. The background model is trained with non-keyword speech as well as background noise and silence. The acoustic modeling schemes for speech units include Gaussian Mixture Model (GMM), Deep Neural Network (DNN), and Time-Delayed Neural Network (TDNN) \cite{sun2017compressed}, and so on. After training, the Vertibi search is applied to find the optimal path in the decoding graph. Whenever the likelihood ratio of the keyword vs. filler model is larger than the pre-defined threshold, the system triggers. 
    
    In recent years, many researchers focus on the DNN based keyword spotting systems, which achieve better performances than traditional methods \cite{ sainath2015convolutional, arik2017convolutional, sun2016max, liu2019loss, wang2017small, shan2018attention, wang2019adversarial, tang2018deep, coucke2019efficient}. In these approaches, a DNN model is trained for words instead of phonemes. The output smoothed posterior probabilities are calculated later from the DNN model's output to compute the confidence score. DNN based methods have the advantages of light-weighting and low latency, which is suitable for real-life applications. As for modeling, many structures based on Convolutional Neural Network (CNN)  \cite{sainath2015convolutional}, Recurrent Neural Network (RNN), Convolutional Recurrent Neural Network (CRNN) \cite{arik2017convolutional}, Long Short Time Memory \cite{sun2016max} (LSTM) and attention mechanism \cite{shan2018attention, wang2019adversarial} are explored. Furthermore,  \cite{tang2018deep} adopts the residual network structure to classify the speech command words, and \cite{coucke2019efficient} introduces a dilated convolutional structure to model the whole keyword sequence, which also shows good performance.
    
    However, in many real-life applications, like smart speakers, the performance of the KWS system is often degraded under the low Signal-to-Noise Ratio (SNR) and far-field conditions. The room reverberation and different kinds of noises in this scenario impose great challenges on the performance of the DNN model, which is trained mainly by close-talking data due to the zero or limit resource for real data collection. A traditional method to tackle this problem is to train DNN models using pooled speech data either collected or simulated from different environments. 
    
    
    In this paper, we employ three domain aware training mechanisms to improve network performance under far-field conditions. The first method is motivated by the noise-robust training with environmental noise embeddings \cite{kim2016environmental, L2013An} in the speech recognition area. We pre-train a domain classifier to extract environmental domain embeddings, which are fused to the training of the keyword classifier. And the second method is inspired by the within-sample variability-invariant loss \cite{cai2020within} and paralleled data training \cite{qian2016investigation, peddinti2016far, li2018developing}  mechanisms successfully applied in speaker verification and automatic speech recognition on complex environments. We propose a training scheme of multi-task learning \cite{caruana_multitask_1997} with the CORAL loss on KWS, which reduces the mismatch of close-talking and far-field conditions in a multi-domain joint learning setup. The third method is based on the multi-task learning \cite{panchapagesan2016multi, giri2015improving}, which optimizes models to make predictions on both domain types and keywords simultaneously. 
    
    This rest of the paper is organized as follows. Section 2 describes the framework of the CNN based KWS system, and in section 3, our proposed domain aware training approaches are introduced. Section 4 discusses the experimental results, and section 5 concludes our work, respectively.

    \section{CNN based KWS system}
    
    
    \begin{figure*}[t]
        \begin{center}
            \includegraphics[width=150mm]{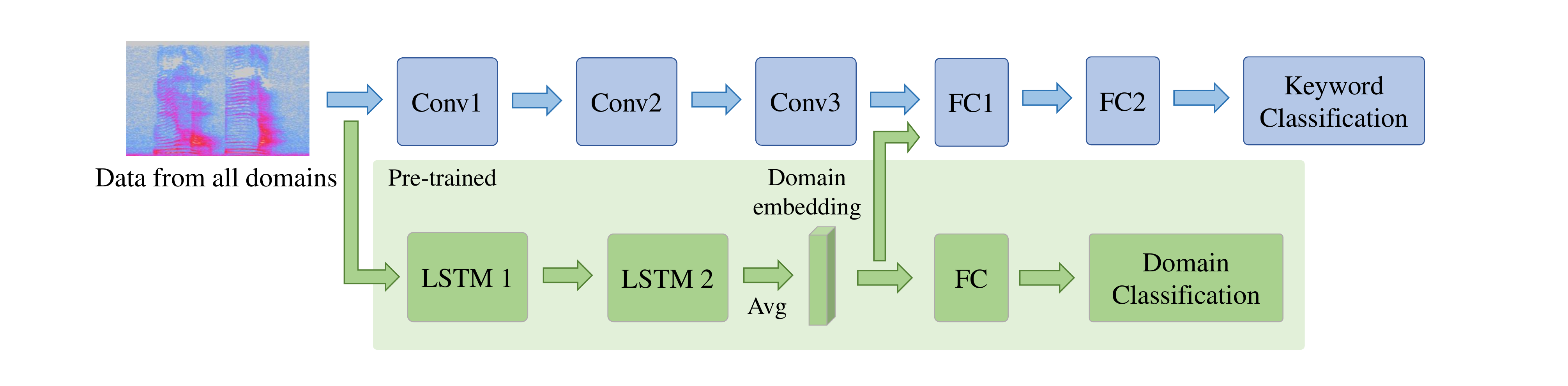}
        \end{center}
        \vspace*{-0.6cm}
        \caption{Framework of the domain embedding system.}
        \vspace*{-3pt}
        \label{ch3_domain_embedding}
        \vspace*{-0.3cm}
    \end{figure*}

    
    Our baseline is constructed on a CNN based KWS system proposed by \cite{sainath2015convolutional}. The pipeline has three main components, feature extraction, network prediction, and confidence computation. In the step of feature extraction,  we extract 40-dimensional log-Mel filterbank energy (Fbank) with a 25ms window and a 10 ms shift. And we apply a window of 40 frames to generate training samples as the input of the model considering the context.
    
    Our convolutional network structure contains three convolutional layers, each of which is followed by a max-pooling layer. The convolutional kernels have the size of $(3, 3)$ with the stride of $(1, 1)$, and the pooling size is set to be $(2, 2)$. Two fully-connected l	ayers accompanied by a final softmax activation layer are then used to predict the target keywords. Rectified linear unit (ReLU) is used as the activation function in hidden layers. 
    
    
    After the training process, the sequence of acoustic features is projected to a posterior probabilities sequence of keywords by the model. In the module of confidence computation, we adopt the method proposed in \cite{liu2019loss, P2015Automatic} to make the decisions. In this approach, we define a sliding window with the length of $T_{s}$ frames which is used to compute scores and denote the input acoustic features in a window as $\bm{x}=\{\bm{x}_1, \bm{x}_2, \cdots \bm{x}_{Ts}\}$. 
    $\bm{w} = \{w_1, w_2 \cdots w_M\}$ represents the words sequence of the pre-defined wake up word. We smooth the output probabilities at a length of $L$ frames by taking average as
    
    \begin{equation}
    s_{w_i}(\bm{x}_t) = \frac{1}{L} \sum_{j=t-L-1}^{t} p_{w_i}(\bm{x}_{j}),
    \end{equation}
    
    where $s_{w_i}(\bm{x}_t)$ represents the smoothed probablities at time $t$ of word $w_i$ and $p_{w_i}(\bm{x}_j)$ refers to the network output of $j^{th}$ frame at word $w_i$. After smoothing, we compute the confidence score as follows:
    \begin{equation}
    h(\bm{x}) = \left [ \max \limits_{1 \leq t_1 < \cdots < t_M \leq T_s} \prod_{i=1}^{M} s_{w_i}(\bm{x}_{ti}) \right ]^{\frac{1}{M}},
    \end{equation}
    where $h(\bm{x})$ refers to the output confidence score. Compared to previous methods \cite{chen2014small}, it has the advantage of considering the order of words that trigger, and at the same time, the time complexity is $O(MT_s)$, which is suitable for the real-time application. The system triggers whenever the confidence score exceed the pre-defined threshold.
    
    \section{Domain Aware Training}
    
    The influence of far-field and noisy conditions in speech signal processing is commonly noticed in many areas like speech recognition and speaker verification. In our works, we apply three domain aware training algorithms on the far-field small-footprint keyword spotting to enrich the knowledge on domains of models. The first algorithm introduces environmental domain embeddings to the keyword classifier. The second method applies correlation alignment loss to reduce the distortion of far-field speech. We also employ multi-task learning to predict keywords and domains simultaneously. 
    
    \subsection{Environmental Domain Embeddings}
    
    In this subsection, we illustrate our approach that optimizes models with environmental domain embeddings derived from a pre-trained domain classifier. This method is inspired by \cite{kim2016environmental}, which explicitly learns the environmental knowledge with the introduction of noise embeddings to the acoustic model. In this paper, we extend this approach to the far-field word-level modeling task.

    Our structure consists of two models: a domain classifier and a keyword classifier. Our domain classifier is optimized with the keyword speech samples recorded from different distances, including 0.25M, 1M, and 3M, which refers to different domain types. The classifier is constructed with a two-layer stacked LSTM structure, followed by an average pooling layer and a final linear layer. Domain embeddings are extracted from the output of the pooling layer. Through this structure, the acoustic features are transformed into a fix-dimensional representation with domain knowledge.
    
    On the base of our CNN model, our keyword classifier is optimized with keyword speech samples and their environmental domain embeddings. Specifically, we extract the acoustic features from the speech and project them to embeddings with the pre-trained domain classifier. And then, the embeddings are concatenated to the output of the penultimate fully-connected layer. The concatenated features are finally fed into a linear layer for the keyword prediction. To further investigate where to concatenate, we also join the embeddings to the output of the last convolutional layer. Figure \ref{ch3_domain_embedding} illustrates the overall architecture.
    
    \subsection{Correlation Alignment}

    \begin{figure*}[ht]
        \begin{center}
            \includegraphics[width=150mm]{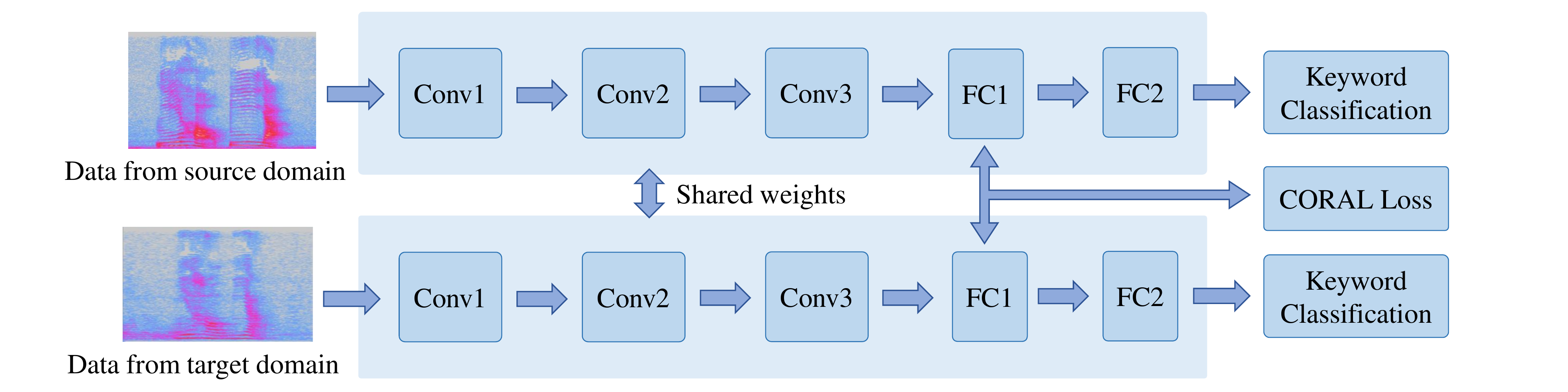}
        \end{center}
        \vspace*{-0.4cm}
        \caption{Framework of the CORAL system.}
        \label{ch3_coral}
        \vspace*{-0.4cm}
    \end{figure*}
    
    The mismatch of inner-class feature distributions on different domains contributes to the degradation of prediction performance. Focusing on this scenario, we apply the CORAL loss to constrain the embedding feature distortions from different domains in the manner of multi-task learning. In our case, we define the penultimate layer of the neural network as our feature layer for alignment loss computation.
    
    CORAL is proposed to align the second-order statistics of the source and target distributions.  \cite{sun2016deep} extend this work to DNN approaches by constructing a differentiable loss functions, which can be used to minimize the distance between outputs of embedding feature layer from different domains.  Suppose the embedding features from source and target domains as $D_S$ and $D_T$. And we denote the dimension of the feature layer as $d$ and the covariance matrices of source and target features are $C_S$ and $C_T$, respectively. The CORAL loss can then be defined as
    \begin{equation}
    l_{CORAL} = \frac{1}{4d^2} \left \|  C_S - C_T \right \|_F^2,
    \end{equation}
    where $\left \| \cdot \right \|_F^2$ denotes the squared matrix Frobenius norm. The covariance matrices of the source and target features \cite{sun2016deep} are
\begin{equation}
C_S = \frac{1}{n_S - 1} (D_S^{\top}D_S-\frac{1}{n_S}(\bm{1}^{\top}D_S)^{\top}(\bm{1}^{\top}D_S)),
\end{equation}
\begin{equation}
C_T = \frac{1}{n_T - 1} (D_T^{\top}D_T-\frac{1}{n_T}(\bm{1}^{\top}D_T)^{\top}(\bm{1}^{\top}D_T)),
\end{equation}
where $n_S$ and $n_T$ represent the number of training samples of source and target domains. $\bm{1}$ refers to a column vector of all 1 elements.
    
    In our work, we compute alignment loss on the outputs of the penultimate layer of the CNN network. Data from three different domains of 0.25m, 1m, and 3m are pooled together for training, and there are several strategies for the loss computation:
    
    \begin{enumerate}
        \item $L = L_{ce} + \lambda L_{coral}(E_{0.25M}, E_{1M})$
        \item $L = L_{ce} + \lambda L_{coral}(E_{0.25M}, E_{3M})$
        \item $L = L_{ce} + \lambda L_{coral}(E_{0.25m}, E_{1M \& 3M})$
        \item $L = L_{ce} + \lambda( L_{coral}(E_{0.25M}, E_{1M}) + \\ L_{coral}(E_{0.25M}, E_{3M})) / 2$
        \item $L = L_{ce} + \lambda (L_{coral}(E_{0.25M}, E_{1M}) + \\ L_{coral}(E_{0.25M}, E_{3M}) + L_{coral}(E_{1M}, E_{3M})) / 3$
    \end{enumerate}

    $\lambda$ is the hyper-parameters representing the weight of alignment loss. The cross-entropy loss $L_{ce}$ is calculated with the logits of data from both the source and target domains. $E_{0.25M}$, $E_{1M}$ and $E_{3M}$ refers to the embedding features used for CORAL calculation. $E_{1M \& 3M}$ means that the $1M$ and $3M$ data are regarded as a group. By minimizing the joint loss, the inner-class embedding feature variabilities between close-talking and far-field domains would be reduced. Figure \ref{ch3_coral} illustrated the whole framework.
    
    \subsection{Multi-task learning}

    \begin{figure}[th]
        \centering
        \includegraphics[width=7cm]{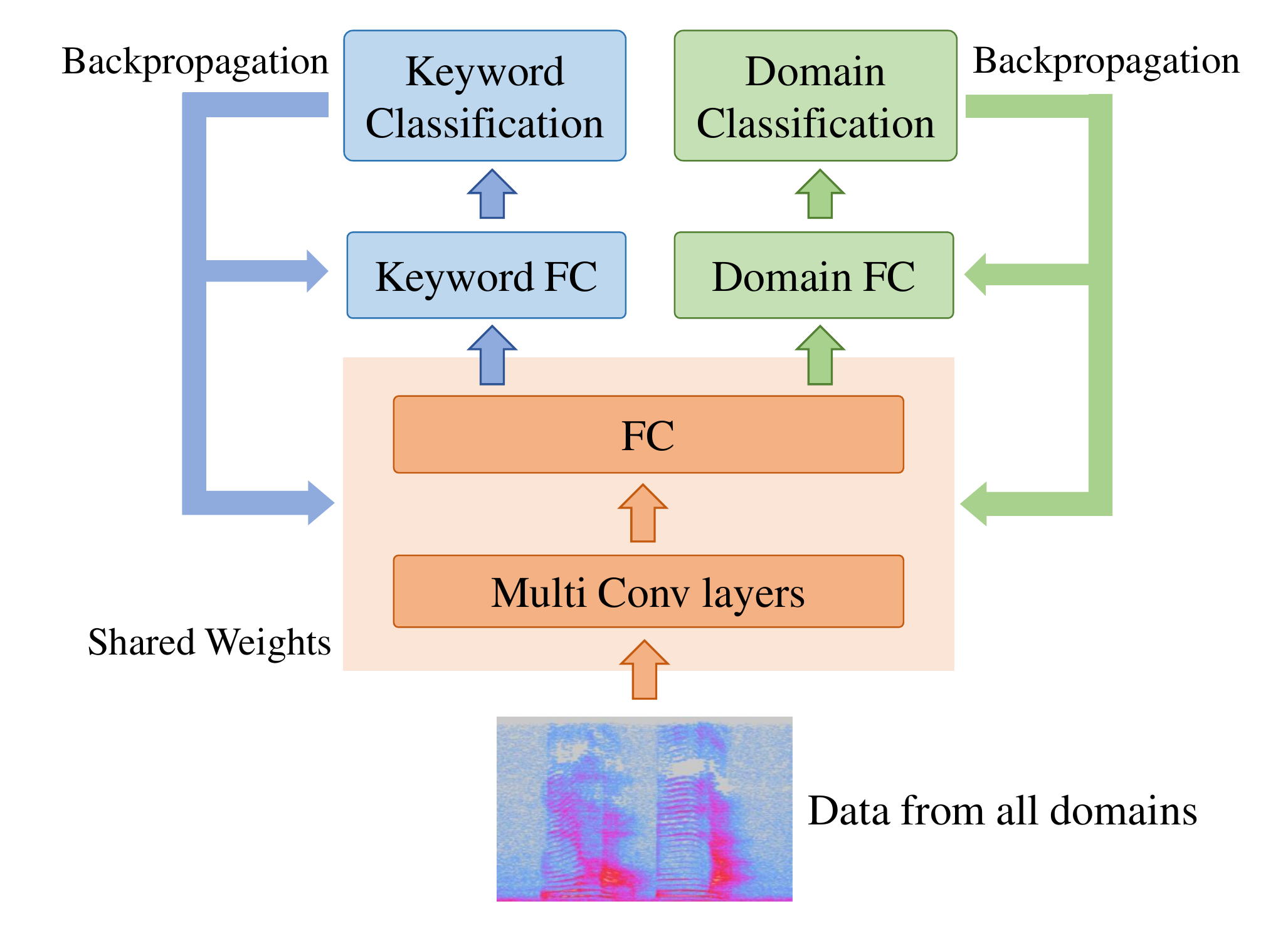}
        \caption{Framework of the MTL system.}
        \label{ch3_mtl_cnn}
    \end{figure}

Multi-task learning (MTL) is a mechanism that simultaneously optimizes the models to learn more than one task with a joint loss function. This method has been successfully applied in many speech-related tasks. In \cite{kim2016environmental}, it is implemented to classify the phonemes and the noise environments to improve the robustness of models toward noisy conditions. Inspired by this work, we perform the MTL algorithm to classify the domains and keywords simultaneously. 

Figure \ref{ch3_mtl_cnn} illustrate our MTL approach. On the base of the baseline CNN structure, an additional fully-connected layer is designed to predict the domain types. The output of the penultimate linear layer serves as a compressed representation with both word and domain information. The previous layers share the weights and are optimized jointly. In the training phase, we calculate the joint cross-entropy with the logits of both the domain and keyword classification. While decoding, only the prediction of keywords is computed.
    
    \section{Experimental results}
    
    \subsection{Data}
    
    Our proposed work is evaluated on a subset of the DMASH dataset \cite{dmash2020xiaoyi}, which is first proposed for the INTERSPEECH 2020 Far-Field Speaker Verification Challenge \cite{dmash2020xiaoyi}. It contains audio of a wake-up word consisting of four Chinese characters, "ni hao, mi ya" ("Hello, Mia" in English) and other sentences that can be utilized as negative data. This dataset includes the speech data recorded by iPhone, Android, microphones, and microphone arrays from various distances. We utilize the recordings of the iPhone from a distance of 0.25m, 1m, and 3m, covering 222 speakers in the training set and 41 speakers in the test set. In our experiment, the 0.25m environment is regarded as close-talking (source domain), and 1m and 3m conditions, are viewed as far-field (target domain). See Table \ref{tab_dataset} for more details of dataset statistics. To objectively measure the data quality, we employ the P.563 algorithm \cite{malfait2006p} on the audio of different distances. Table \ref{tab_snr} illustrate the mean opinion score (MOS) results.

     \begin{table}
         \centering
         \caption{MOS of data from different domains}
         \begin{adjustbox}{max width=0.99 \columnwidth}
             \begin{tabular}{c c c c c}
                 \toprule
                 & 0.25m & 1m  &  3m \\ \midrule
                 MOS & 2.698 & 2.022 & 1.375   \\
                 \bottomrule
             \end{tabular}
         \end{adjustbox}
         \label{tab_snr}
     \end{table}
    
    \begin{table}[]
        \centering
        \caption{Dataset statistics.}
        \vspace{-0.2cm}
        \begin{adjustbox}{max width=1 \columnwidth}
            \begin{tabular}{lcccc}
                \toprule
                &               & utterances &  positive  & negative \\ \midrule
                & 0.25m    & 178k  & 19k  & 159k  \\
                Train     & 1m & 146k   & 15k  & 131k    \\
                & 3m & 143k  & 15k  & 128k   \\ \midrule
                & 0.25m         & 37k & 4k & 33k \\
                Evaluation        & 1m & 32k       & 4k  & 28k \\
                & 3m & 31k & 4k & 27k \\ \bottomrule
            \end{tabular}
        \end{adjustbox}
        \label{tab_dataset}
        \vspace*{-0.6cm}
    \end{table}
    
    \subsection{Experiment setup}
    
    We determine target word labels by force-alignment with an LVCSR system trained with the AISHELL-2 dataset \cite{du2018aishell}. Here, for keyword "ni hao, mi ya", we find out the ending time of "ni", "hao", and "mi", and include its previous 20 frames and next 20 frames to construct a window of 40 frames. Log fbank is adopted as our input acoustic features. The baseline system is trained with cross-entropy loss. Stochastic gradient descent with Nesterov momentum is selected as the optimizer. The learning rate is first initialized as 0.01 and decreases by a factor of 0.1 when the training loss plateau. We train the CNN model for 100 epochs with a batch size of 128 and employ early stopping when the training loss is not decreasing. In the evaluation period, we use a sliding window of 100 frames to compute the confidence score.
    
    As the baseline system, we pool data from both close-talking and far-field conditions for training. In our experiments, for deep CORAL training, we set the weight $\lambda$ to 0.2, 0.4, 0.6, 0.8, and 1.0, respectively. In our preliminary experiments, we find out that 0.8 is a suitable parameter, so our experiments on the CORAL loss are done under this weight. For MTL training, after the preliminary experiments with $\lambda=\{0.1, 0.2, 0.3, 0.4, 0.5\}$, we observe that the system achieves the best overall performance when $\lambda=0.2$.
    
    The performance is measured with the false reject (FR) rate under one false alarm (FA) per hour. 
    \subsection{Results}

\begin{table}[h]
    \centering
    \caption{Performance of the baseline system (the false reject (FR) rate (\%) under one false alarm (FA) per hour)}        
    \label{result_baseline}
    \begin{tabular}{cccccc}
        \toprule
        Training set & 0.25M & 1M & 3M \\
        \midrule
        Only 0.25M    & 1.29    & 2.91    & 11.6 \\
        Only 1M     & 2.03    & 1.58 & 7.77 \\
        Only 3M & 10.9    & 8.00 & 10.6 \\
        Mix of 0.25M and 1M & \textbf{0.91}    & \textbf{1.38} & 6.06\\
        Mix of 0.25M and 3M & 1.54    & 1.97 & \textbf{5.60}\\
        Mix of all distances& 1.41    & 1.64 & 6.33 \\
        \bottomrule
    \end{tabular}
    \vspace*{-0.4cm}
\end{table}

The performance of the baseline system is illustrated in Table \ref{result_baseline}. From the results, we can obtain the following observations. First, with the increase of recording distance, the distortion becomes severer, and the performance of the baseline system degrades. Second, for the 0.25M and 1M datasets, when the training set and test set are from the same domain, the system performs better than the scenarios of domain mismatch. The network trained with only 3M datasets shows poor performance in every test set. Third, pooling the close-talking domain and target domain training data helps improve the performance on the target domain's test set. And the performance of the close-talking condition can still be maintained. The system trained with data from all fields has a balanced performance, while it is worse than the models trained with its target domains.

\begin{table}[h]
    \centering
    \caption{Performances of models trained with different methods on the test sets}
    \vspace{-0.3cm}
    \label{results_all}
    \begin{tabular}{@{\ \ }c@{\ \ \ \ \ \ \ \ \ \ }c@{\ \ \ \ \ \ \ \ \ \ }c@{\ \ \ \ \ \ \ \ \ \ }c@{\ \ }}
        \toprule
        Model name  & 0.25M & 1M & 3M \\
        \midrule
        EMB1   & 1.11    & 1.59    & 4.99 \\
        EMB2   & 1.21    & \textbf{1.02}    & \textbf{4.11} \\
        \midrule
        CORAL1 & 1.37    & 1.05    & 4.69 \\
        CORAL2 & 1.19    & 1.41    & 5.02 \\
        CORAL3 & \textbf{1.09}    & 1.52    & 5.97 \\
        CORAL4 & 1.27    & 1.47    & 5.21 \\
        CORAL5 & 1.21    & 1.41    & 4.78 \\
        \midrule
        MTL   & 1.70    & 1.44    & 5.15 \\
        \bottomrule
    \end{tabular}
    \vspace*{-0.6cm}
\end{table}

The results of our proposed system are shown in Table \ref{results_all}. The EMB1 system represents the system that concatenates domain embeddings to the output of the penultimate linear layer. And the EMB2 system concatenates to the output of the last convolutional layer. The CORAL1 to CORAL5 systems denotes the five different CORAL calculating approaches described in section 3. From the table, we can have the following findings. 1) In the domain embedding approaches, the EMB2 system outperforms the EMB1 system on the far-field conditions. The concatenation of embeddings in an early stage of the network helps the network better learn the domain information from the embeddings. 2) Among the CORAL systems, the CORAL1 system produces the best results on the far-field conditions. The CORAL2 system obtains worse scores than the CORAL1 system, which indicates that this method is sensitive to the domain types. From the results of the CORAL3 and CORAL4, we can see that regarding 1M and 3M datasets as a group are unhelpful to classification. The CORAL5 system calculates the CORAL loss for each pair of domains and achieves balanced results. 3) The MTL system obtains satisfying improvement on far-field speech while it has a relatively large degradation on the close-talking set.

Comparing different algorithms, we can find that systems based on domain embeddings achieve the best improvement, and the CORAL systems also outperform the baseline system on the far-field conditions. The CORAL system has the advantage that it does not require any extra network structures. The domain embedding system has an additional domain classifier, which increases the number of network parameters and the complexity of decoding computation. The MTL method is not as effective as the other two approaches.
    
    \section{Conclusions}
    
    In this paper, we concentrate on the task of small-footprint keyword spotting under the far-field environment. Far-field environments are commonly noticed in real-life speech applications, and it causes serve degradation of performance due to room reverberation and various kinds of noises. To cope with the distortions, we employ three domain aware training schemes, including learning with domain embeddings, with the CORAL loss, and MTL with inputs from different domains of data. Experimental results show that our methods manage to maintain the performance on the close-talking test dataset and achieve significant improvement in far-field conditions. Approaches with domain embeddings deliver the best performance while increasing the model size and computing cost. The CORAL systems also outperform the baseline system without changing the model structure, while it is sensitive to the domain types of data. The MTL approach is less effective than the other two methods.
    
    \section{Acknowledgment}
    This research was funded by Kunshan Government Research (KGR) Funding in AY 2019/2020.
    
    \vfill
    \pagebreak
    
    \bibliographystyle{IEEEtrans}
    \bibliography{mybib}
    
\end{document}